\documentclass[a4paper,onecolumn,aps,prl]{revtex4}
\usepackage{graphicx}
\usepackage{amsmath}
\usepackage{color}
\usepackage{cancel}
\usepackage{ulem} 
\usepackage{marginnote}
\usepackage{tikz}
\newcommand{\ket}[1]{\ensuremath{\left|{#1}\right\rangle}}
\newcommand{\bra}[1]{\ensuremath{\left\langle{#1}\right|}}

\begin{document}

\title{Impurities as a quantum thermometer for a Bose-Einstein condensate}

\author{Carlos Sab{\'\i}n$^{1}$, Angela White $^{2}$, Lucia Hackermuller $^{3}$,  Ivette Fuentes$^{1}$ }

\affiliation{
$^{1}$School of Mathematical Sciences, University of Nottingham, University Park,
Nottingham NG7 2RD, United Kingdom\\ $^{2}$ Quantum Systems Unit, Okinawa Institute of Science and Technology Graduate University, Onna-son, Okinawa 904-0495, Japan\\$^{3}$School of Physics and Astronomy, University of Nottingham, University Park,
Nottingham NG7 2RD, United Kingdom\\
$^*$ Corresponding author: \texttt{carlos.sabin@nottingham.ac.uk}}
%

\begin{abstract}
{\bf We introduce a primary thermometer which measures the temperature of a Bose-Einstein Condensate in the sub-nK regime. We show, using quantum Fisher information, that the precision of our technique improves the state-of-the-art in thermometry in the sub-nK regime. The temperature of the condensate is mapped onto the quantum phase of an atomic dot that interacts with the system for short times. We show that the highest precision is achieved when the phase is dynamical rather than geometric and when it is detected through Ramsey interferometry.  Standard techniques to determine the temperature of a condensate involve an indirect estimation through mean particle velocities made after releasing the condensate.  In contrast to these destructive measurements, our method involves a negligible disturbance of the system. }
\end{abstract}
\maketitle
Temperature is a crucial concept in quantum physics. Paradigmatic phenomena such as superconductivity and Bose-Einstein condensation only occur below a critical temperature.  Bose Einstein Condensates (BECs) \cite{pethicksmith} allow the study of quantum effects in systems consisting of up to $10^{8}$ atoms by cooling them to regimes in which the individual atomic wavefunctions overlap.  In this case, the system exhibits  quantum behaviour at mesoscopic scales.  

\begin{figure*}[t] 
\includegraphics[width=\textwidth]{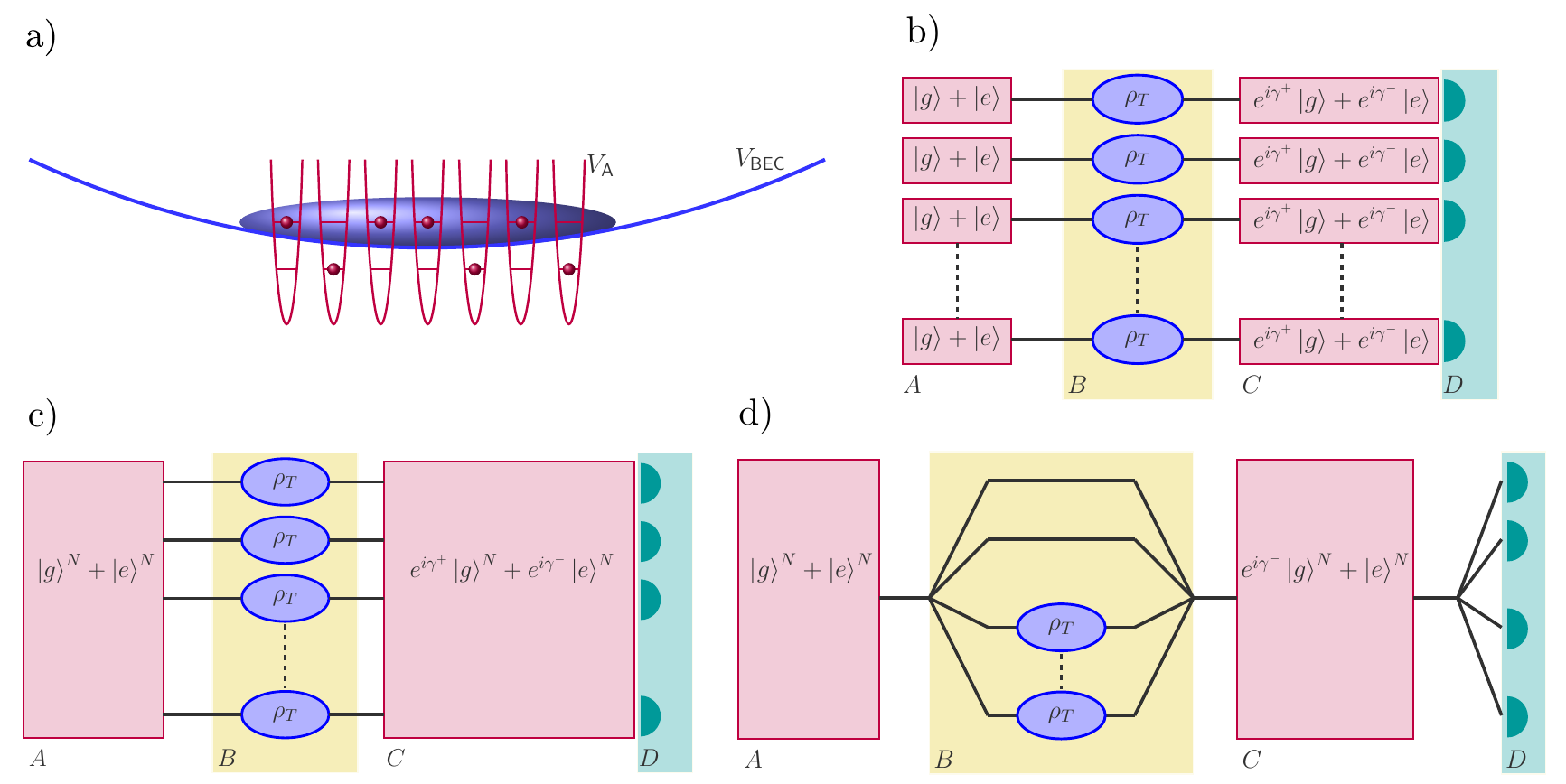}
\caption{a) Sketch of the experimental setup: several atomic quantum dots are embedded within a BEC reservoir. They are coupled at different times through a Raman transition to the phononic fluctuations of the BEC. The use of many dots allows one to implement many measurements and improve the accuracy.   b) Ramsey interferometry scheme to measure the relative dynamical phase and
hence measure temperature. c) Ramsey interferometry scheme with entangled input states.  Entanglement can in priciple improve the precision, reaching the Heisenberg limit. d) Mach-Zehnder interferometer scheme with entangled input states. A. Input state, B. Interaction switched on, C. Final state D.Readout }
     \label{fig:fig1}
\end{figure*}

Very low temperatures are also required to observe quantum field theory effects e.g.  Unruh-Hawking radiation \cite{papermort}.
 Quantum field theory predicts that the dynamics of spacetime or the presence of horizons produce quantum particles from vacuum fluctuations.  For example, in the dynamical Casimir effect, a moving boundary condition gives rise to vacuum excitations. This effect was recently demonstrated in superconducting circuits  \cite{casimirwilson}. Currently several experimental groups in the field of analogue gravity  attempt to demonstrate cosmological particle creation, dynamical Casimir effect and Hawking radiation in analogue spacetimes produced in BECs \cite{casimirwestbrook}.  However,  in a BEC a thermal background is always present due to unavoidable atomic collisions. In these experiments, it is crucial to work at low temperatures so that the quantum particles created by effects of emergent spacetimes can be distinguished from this thermal noise. Recent experiments to demonstrate the dynamical Casimir effect in a  BEC \cite{casimirwestbrook}, have not been able to ensure that the excitations produced are indeed quantum even at temperatures as low as 200 nK.
A new method to accurately determine temperatures in the nK and sub-nK regime would be of great benefit to these experiments. 

The temperature of a BEC is commonly estimated indirectly by comparing the density profile of the atoms  with a velocity distribution \cite{pethicksmith,reviewketterle,apsortion, apsortion2}. This density profile is determined by absorption imaging which is a destructive method that involves releasing the condensate. Although accuracies of $\sim 1 \%$ can be achieved in the 100 $nK$ regime \cite{temps100}, the relative error in the measurement of temperatures considerably grows at very low temperatures. The best accuracies that have been reported in the $\operatorname{nK}$  and sub-$\operatorname{nK}$ regime are of $\sim20 \%$ \cite{sciencewr}.  Several alternative methods to measure the temperature of a BEC have been proposed. For instance, phase-contrast imaging  \cite{phasecontrast} can be used to determine the temperature through the phase shift of a probe laser beam that interacts with the sample. However, at temperatures of a few $nK$ and below this method has low spatial resolution and can only be applied reliably if the condensate is allowed to freely expand for a given time \cite{stingerjournallowT}. Another example is noise thermometry with two coupled Bose-Einstein condensates \cite{gati}, which has been implemented in a regime in which quantum fluctuations are small ($50-80 \operatorname{nK}$). Therefore, absorption imaging is the standard technique for measuring temperatures in the nK regime \cite{sciencewr} and, in general, for probing the condensate \cite{sciencemicrog}. 

Recent theoretical and experimental developments in cold atom gases have lead to the experimental demonstration of systems consisting of mixtures of two different atomic species or two hyperfine states of the same species \cite{scienceKRb,prlKRb,prlmixturebosefermi,feshbach1,feshbach2}.  In particular, it is in principle possible to create an atomic quantum dot by trapping  a few atoms in a tight trap provided by a laser beam \cite{recati,quantumtweezer,atomchip,subpois,subpois2,singleatombec,singleimpurities}.  The number of particles in the dot is determined by the interatomic interactions which can be modulated experimentally, through a Feshbach resonance \cite{feshbach1,feshbach2,feshbach3}. In this paper we show how an atomic quantum dot immersed within the condensate can be used to measure the temperature of the condensate with high precision and a negligible disturbance of the system.  The scheme can be implemented using an optical lattice with a single atom per site \cite{meschede}, where each atom interacts at a different time with the condensate. This system constitutes a primary thermometer which measures the temperatures ranging from the sub nK regime to the condensate critical temperature. 

The outline of the paper is as follows. We start by showing how a two-level system interacting with a quantum field in a thermal state through the standard Jaynes-Cummings Hamiltonian can acquire a dynamical phase that depends on the temperature of the field. This phase is then read out through Ramsey interferometry. We use the Quantum Fisher Information to quantify the accuracy in the temperature estimation. We show that with realistic experimental parameters, the precision of our thermometer improves on state of the art thermometry in the nK and sub nK regimes already after a small number of measurements. Since each dot embedded within the condensate provides one measurement, it is important to ensure the statistical independence of the individual measurements, by making the dots interact independently with the condensate. By preparing the dots in an entangled state, it is possible to further improve the precision and reach the Heisenberg limit \cite{esteve}, that is the limit imposed fundamentally by quantum theory. Finally, we present details of possible physical implementations of the thermometer using atomic quantum dots coupled to phonons in a BEC (see Fig.\ref{fig:fig1}). 

While the idea of probing several aspects of the condensate with different two-level systems  and interactions has been introduced in earlier works (see, for instance, \cite{ngbosetwolevelBEC,bruderer}) our work represents a step forward in several directions, which we outline here. We show that with realistic experimental parameters the system can be described by a Jaynes-Cummings model, where the dots are effectively coupled to a single mode of the quantum phase fluctuations of the condensate. Moreover, we show that is possible to achieve a regime in which spontaneous transitions of the dots are negligible (see Appendix A) and thus the time evolution of the system is given simply by dynamical and geometrical phases.  We study both phases and analyse the advantages of using Ramsey vs. Mach-Zehnder interferometry to estimate relative phases. We use Quantum Metrology techniques to quantify the precision in each scenario. We show that the best precision is achieved when using dynamical phases and Ramsey interferometry. Furthermore, we show that this precision can improve on standard methods to determine temperature in the nK and sub nK regimes.

\section*{Results}
We now proceed to introduce our methods and results.
We consider the Jaynes-Cummings model that describes the interaction of a two-level system of frequency gap $\Omega_d$ and one mode of a quantum field with frequency $\Omega_a$ 
\begin{equation}
H_{JC} (t)= \hbar\,\Omega_a a^\dagger a+\hbar\,\frac{\Omega_d}{2} \sigma_z +\hbar g (\sigma^-a^\dagger e^{i\theta}+\sigma^+ a e^{-i\theta}), \label{eq:hamiltonianJC}
\end{equation}
where $a$, $a^{\dagger}$ are creation and annihilation field operators, $\hbar=h/(2\pi)$ is the reduced  Planck constant, $\sigma^{\pm}=(1/2)(\sigma_x\pm i \sigma_y)$ and $\sigma_x$ and $\sigma_y$ are Pauli matrices. The coupling strength between the field and atom is given by $g$, the detuning is $\delta= |\Omega_a-\Omega_d|$ and the phase $\theta=k\,x-\delta t$ is a function of time and the position $x$ of the atom. As we will explain in more detail later, with suitable boundary conditions an atomic dot interacting with the phonon field of a BEC can be described with this simple interaction in the $nK$ and sub-$nK$ regime. 

\begin{figure*}[t]
\hspace*{-0.4cm}
\includegraphics[width=\linewidth]{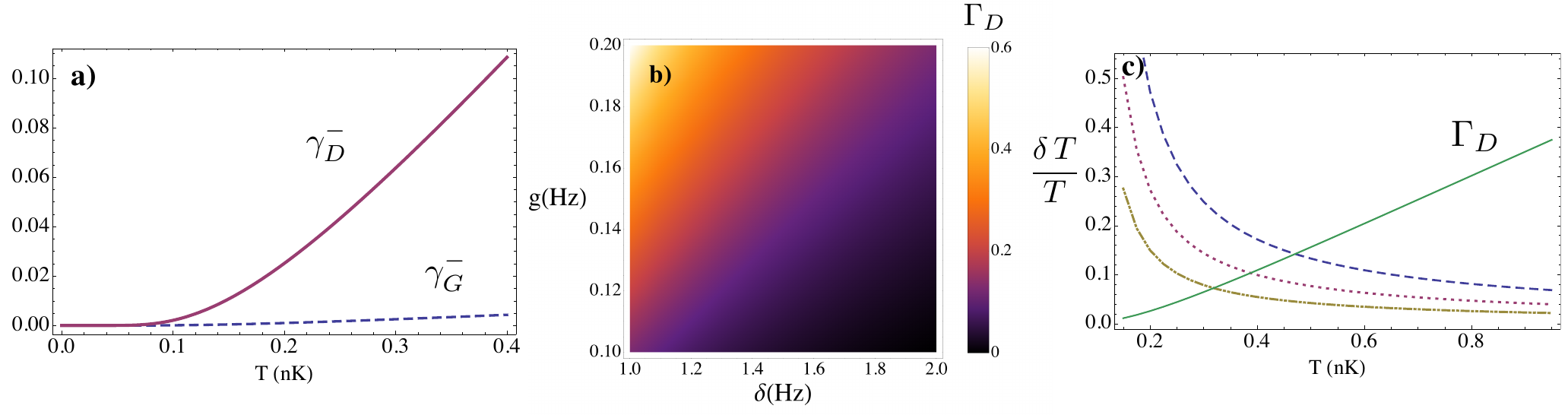}
\caption{ a) Comparison of geometrical (dahed, blue) $\gamma_G^-$ vs. dynamical  $\gamma_D^-$ (solid, red) with  $\Omega_a= 2\,\pi\times10\,\operatorname{Hz}$,  $g= 2\,\pi \times 0.2\,\operatorname{Hz}$, $\delta= 2\,\pi \times 2\,\operatorname{Hz}$ and $c=5\, \operatorname{mm}/\operatorname{s}$. $\gamma_D^-\simeq 8\pi\,\gamma_G^-$. The dynamical phase is much more sensitive to the temperature in the sub-$nK$ regime. b) Dependence of the dynamical phase $\Gamma_D$ on $g$ and $\delta$ at $T=0.5\,\operatorname{nK}$. $\Gamma_D$ is sensitive to the ratio $g/\delta$. c) $\Gamma_D$ (green, solid) and relative error $\delta T/T$ after 1000 (blue, dashed), 3000 (red, dotted) and 10000 (yellow,dash-dotted) measurements. The parameters are the same as in a). The precision in the best case is around 1$\%$ and can be improved by increasing the number of measurements and using entangled input states.}\label{fig:results}
\end{figure*}

The temperature of the field will be determined through the phases acquired by the atom through its evolution under the Hamiltonian (\ref{eq:hamiltonianJC}), assuming the adiabatic approximation (see Methods). 
The total accumulated phase $\gamma$ can be split into a geometric and a dynamical part. It has been shown that the geometric phase can be used to gain information about the quantum state of a bosonic scalar field. For pure field states, the phase encodes information 
about the number of particles in the field \cite{ivyvlatkoberry}. In particular, for initially
squeezed states, the phase depends also on the squeezing strength \cite{ivytraps}. More recently, the geometric phase was employed to estimate  the temperature of a field inside a cavity \cite{thermoeduivy}. We will apply Quantum Fisher Information to show here that estimating the dynamical phase is the optimal way for measuring temperatures.
We consider the case where the field is initially in a thermal state $\rho_F=\sum_n p_n\ket{n}\bra{n}$ with temperature $T$. The probability distribution is given by $p_n=(e^{-F})^n\,(1-e^{-F})$ where $F=\frac{\hbar\Omega_a}{k_B\,T}$, and $k_B$ is Boltzmann's constant.  Here we have used the number state basis spanned by $\ket{n}=(a^\dagger)^n\ket{0}$, where $n$ is the number of phonons of frequency $\Omega_a$.  We find (see Methods), that the geometric phase is 
\begin{equation}
\gamma_G^{-}\simeq \frac{g^2}{\delta^2(e^F-1)} \label{eq:geometricalappr}
\end{equation}
and the dynamical phase 
\begin{equation}
\gamma_D^{-}\simeq \frac{4 t g^2}{\delta(e^F-1)}. \label{eq:dynamicalappr}
\end{equation}
As explained in Methods, in order to compare the phases we consider $t=2\pi/\delta$, since cyclic evolution is necessary to compute the geometric phase given in Eq. (\ref{eq:geometricalappr}). We find that the dynamical phase is always larger than its geometric counterpart 
 \begin{equation}
 \gamma_G^{-}\simeq\frac{\gamma_D^{-}}{8\pi}. \label{eq:geometricvsdynamical}
 \end{equation}
  In Fig.\ref{fig:results}a) we plot the dynamical and geometric phases for realistic experimental parameters.  One observes that for temperatures in the sub nK regime, the geometric phase is negligible in comparison to the dynamical phase. Furthermore, we will show that the precision obtained through the dynamical phase in the measurement of the temperature is higher than the precision attainable with the geometric one.

Since the phases acquired by the system are global and cannot be measured directly, it is necessary to setup an interferometric scheme that will allow us to measure relative phases. In particular, the geometric phase can only be detected using a Mach-Zehnder interferometer.  In this type of interferometer the qubit is split into two different trajectories and is made to interact with a thermal field in one arm of the interferometer acquiring a relative phase that can be measured at the interferometer's output. Now we will use standard techniques of phase estimation in Quantum Metrology to estimate the precision of the temperature measurement in this experiment (see Methods) and determine which type of phase (geometrical or dynamical) is more convenient. 

We consider that in one arm of the Mach-Zehnder interferometer the qubit, initially in state $|g\rangle$, interacts with a field in state $\rho_T$. In the second arm we assume that the qubit in state  $|e\rangle$ does not interact with the field. Since, up to the first order in $g/\delta$ the state $\ket{g}\bra{g}\otimes\rho_T\simeq\sum P_n \ket{n-}\bra{n-}$, the relative phase at the output will therefore be given by $\gamma^{-}=\gamma_D^{-}+\gamma_G^{-}$. For the sake of comparison, we first assume that the dynamical phase can be cancelled \cite{spinecho} and we calculate the quantum Fisher information to estimate the precision on the temperature measurement through the geometrical phase. Since in this case $F(\rho_T)=(\partial_T\gamma_G^-)^2$, the Cramer-Rao bound in the geometric case employing Mach-Zehnder interferometry, denoted by $\delta T_M^{G}$, yields
\begin{equation} 
\delta T_M^{G}\geq\frac{1}{\sqrt{M}\,\partial_T\gamma_G^-}>\frac{T\,\delta^2\,(e^F -2)}{\sqrt{M} F\,g^2}.\label{eq:geometricM}
\end{equation}
Here the bound is obtained using a series expansion in $g/\delta$. 
Following the same procedure we find bounds to the quantum Fisher information for the case where the temperature is measured using the dynamical phase. We find that:
\begin{equation}  
\delta T^D_M\leq\delta_0;\,\frac{T\,\delta^2\,(e^F -1)^2}{\sqrt{M} 8\,\pi\,F\,g^2\,e^F}<\delta_0<\frac{T\,\delta^2\,(e^F -1)^2}{\sqrt{M} 4\,\pi\,F\,g^2\,e^F}.\label{eq:dynamicM}
\end{equation}
From Eq. (\ref{eq:geometricM}) and (\ref{eq:dynamicM}) it follows that the error in estimating the temperature using the dynamical phase is smaller $\delta T^D_M<\delta T^G_M$. Furthermore, using the dynamical phase allows us to simplify the setup by using Ramsey rather than Mach-Zehnder interferometry. Ramsey interferometry does not involve spatial splitting the atomic wave function to undergo different trajectories and therefore is less demanding from the experimental viewpoint.

In Ramsey interferometry the two-level system is prepared in a superposition state, $\ket{\uparrow}=\ket{g}+\ket{e}$,  and  is allowed to interact with the quantum field in a thermal state $\rho_T$ during a time $t$.  After the interaction, the probability of finding the two-level system in the excited state depends on the dynamical relative phase, $\Gamma_D$, picked up by it. Under weak adiabatic evolution the state $\ket{\uparrow}\bra{\uparrow}\otimes\rho_T\simeq\sum P_n (\ket{n-}+\ket{n+})(\bra{n-}+\bra{n+})$ acquires the relative phase, $\Gamma_D=\gamma_D^+-\gamma_D^-$, which is very sensitive to the temperature in an ultralow regime such as nK and sub-nK. Here, $\gamma_D^+$, $\gamma_D^-$ are the dynamical phases acquired by $\ket{+}$ and $\ket{-}$ respectively (see Appendix A). The Fisher information in this case is given by $F(\rho_T)=(\partial_T\gamma_D^+ -\partial_T\gamma_D^-)^2=4(\partial_T\gamma_D^-)^2$. Therefore, the achievable precision in the temperature measurement is increased by a factor of 2 for
Ramsey interferometry, i.e. $\delta T^D_R=(1/2)\delta T^D_M$. 

In Fig. \ref{fig:results}b we plot the dependence of $\Gamma_D$ on the detuning $\delta$ and the coupling strength $g$ for the relevant experimental regime that we consider below. Note that the phase is sensitive to $g/\delta$ and is essentially determined by  this ratio and $\Omega_a$. In Fig. \ref{fig:results}c we plot the relative error $\delta T^D_R/T$ after 1000, 3000 and 5000 measurements. In the best scenario, the relative error achieved is around 1 $\%$ in the range $0.3-1\operatorname{nK}$, which improves the 20 $\%$ of the current state of the art thermometry in the sub $\operatorname{nK}$ regime \cite{sciencewr}. 

\subsection{Experimental implementation} 
We now consider a realistic implementation of the above results. We will consider a system consisting of a BEC superfluid reservoir in a shallow confining trap interacting with an array of atomic quantum dots \cite{recati,Orth2008}.  Each dot is created by applying a localised steep potential that traps atoms of a different hyperfine state of the same atomic species as the BEC.  As we will see below this system obeys the Jaynes-Cummings Hamiltonian where the coupling is mediated by a Raman transition. We would like to point out that two-level systems can also be implemented using a deep double-well optical lattice loaded with cold atoms in a single-occupancy regime \cite{recatiII}. This system also obeys the JC Hamiltonian. However, in this paper we will discuss in more detail the atomic dot implementation \cite{recati,Orth2008}. 

In the absence of atomic collisions, the BEC can in principle reach absolute zero temperature. Under this assumption, the condensate is commonly described by a classical density function. In realistic scenarios, collisions are always present and therefore, in the superfluid regime, the condensate is better described by a mean field classical background plus quantum fluctuations. The fluctuations, for frequencies lower than the so-called healing length,  are given by a phononic quantum field operator $\Pi(\bf{x})$, which can be expanded in terms of Bogoliubov modes. Now we proceed to describe the quantum dots and their interaction with the phononic field. Each dot is created by applying a localised steep potential which traps atoms of a different hyperfine state of the same atomic species as the BEC. Alternatively, in \cite{recatiII} the two-level systems are created by loading cold atoms of a different species in an optical lattice consisting of double well potentials where only single occupancy of each site is allowed. Focusing on the former, we choose a large collisional interaction strength $g_{aa}$ inside each dot well, to guarantee  that the occupation number in each site is either $0$ or $1$, giving rise to an array of two level systems. We also assume that the wells are deep and separated enough to neglect direct interaction between different sites \cite{Orth2008}. 

The atoms in the BEC and the dots can be coupled through a Raman transition using external lasers, giving rise to a phonon-mediated interaction \cite{recati}. The effective Hamiltonian obtained under the above assumptions can be written as \cite{recati,Orth2008}: $H= H_A+ H_B+H_{AB}$. Here $H_A$ is the free Hamiltonian of an array of $M$ dots, which is given by  $H_{A}=\sum^{M}_{i=1} \hbar \Omega_d \sigma_z^{i}/2$. The frequency $\Omega_d$ is a function of the effective Rabi frequency  of the Raman transition $\Omega$ and the number of atoms in the condensate \cite{recati}. $H_B$ is the free Hamiltonian for the phonons $H_B=\sum_k \hbar\omega_k a^{\dagger}_ka_k$, where $\omega_k=c\,|\bf{k}|$ and $c$ is the speed of sound. Finally, the interaction Hamiltonian is given by:
\begin{equation}
H_{AB}=-\hbar\,\delta'\sum^{M}_{i=1} \sigma^i_z+\hbar (g_{ab}-g_{bb})\sum^{M}_{i=1} \Pi({\bf{x^i}})\sigma^{i}_x,\label{eq:interaction}
\end{equation}
where $\bf{x^i}$ is the position of the dot in the condensate. The coupling strengths between dot-condensate and condensate-condensate atoms are given by $g_{ab}$ and $g_{bb}$, respectively. The frequency $\delta'$ is a function of $g_{ab}, g_{aa}$, the effective Rabi frequency and the detuning. We assume that all the coupling strengths $g_{ab}, g_{aa}$ and $g_{bb}$ can be tuned with Feshbach resonances. In particular, we will consider that the strengths are tuned such that $\delta'=0$. In this case the system is described by the spin-boson Hamiltonian \cite{recati}, in which the structure of the reservoir is characterised by the spectral density. The ubiquity of physical systems that can be described by the spin-boson Hamiltonian directly illustrates that the thermometry technique we describe can be applied to a range of different physical systems.  

With suitable boundary conditions for the condensate trap -e. g. hard-wall or box potentials \cite{condensatebox1,condensatebox2}-, the energy separation of the phononic modes can be large enough to ensure that each two-level system is only effectively coupled to the mode with closest frequency to $\Omega_d$.  We denote this frequency by $\Omega_a$ and apply the Rotating wave Approximation (RWA). Notice however that we consider the JC model only for the sake of mathematical simplicity. Indeed, if the interaction is given by the more general spin-boson model (Eq. (\ref{eq:interaction})) our results can be extended by using perturbation theory techniques with the multimode field.
Assuming that each dot interacts with the condensate at different times, the Hamiltonian for each dot is given by Eq. (\ref{eq:hamiltonianJC}), where now the dot-field coupling strength $g$ is given by:   
\begin{equation}
g=\sqrt{\frac{c\,k}{2 \hbar Vg_{bb}}} (g_{ab}-g_{bb}),\label{eq:couplingBEC}
\end{equation}
where $V$ is the condensate volume. The RWA holds for weak couplings and small detunings $\{g,\delta\}<< \epsilon$ where $\epsilon=\Omega_a+ \Omega_d$. Taking typical values for the condensate length $L= 500\operatorname{\mu m}$ and the speed of sound $c=5 \operatorname{mm}/s$ and assuming $\lambda_f\simeq L$, the frequency of the phonons is given by $\Omega_a=2\pi\times10\,\operatorname{Hz}$. With experimental data for the Feshbach resonances \cite{feshbach1,feshbach2,feshbach3} the values of $g$ can be tuned within a broad range from $2 \pi\times0.1\operatorname{Hz}$ to $2 \pi\times\operatorname{10}\operatorname{Hz}$. Taking $g$ around $0.1\, \operatorname{Hz}$ and a detuning of $2\pi\times1-2\,\operatorname{Hz}$ we are consistent with both RWA and the adiabatic approximation. This is the range of parameters explored in the plots. 
 
Dissipation in this model is characterized by \cite{recati} $\alpha=(\frac{\rho_b\,  g_{ab}}{m c^2 }-1)^2=(\frac{g_{ab}}{g_{bb}} -1)^2$. As can be seen in Eq. \ref{eq:couplingBEC}, the weak coupling regime that we are considering entails that $g_{ab}$ is slightly larger than $g_{bb}$. In particular, for $g=0.1 \operatorname{Hz}$, $g_{ab}/g_{bb} =1+1/70$ and then $\alpha= 2\cdot10^{-4}$. Therefore, we are in the regime $\alpha<<1$, where the damping rate $\Gamma$ of the Rabi oscillations is negligible as compared with the frequency of the dot $\Omega_{d}$, and the dynamics is well described by perfect Rabi oscillations, as in the Jaynes-Cummings model.
 
 The description of the experiment is the following. The atomic dots are prepared in a separable superposition of their internal levels,
$\otimes_{i=1...M}\ket{\uparrow_i}$. They are then coupled to the phonons through Raman transitions and allowed to interact with the BEC for some time, which in our plots is of the order of  0.1$s$. The times can be made significantly shorter by relaxing the cyclicity condition, which we imposed in order to compare the dynamical phase analysis with the
geometrical phase analysis. However, it is no longer necessary in this setting.  The probabilities of excitation of all the dots are then measured using standard imaging techniques \cite{meschede}. In order to ensure that the measurements are completely independent, one possibility is that each dot interact at different times with the condensate. Moreover, the separation in time between interactions should be large enough to rule out temporal correlations between the occupation numbers of the considered mode. These temporal correlations drop to 0 after a few $\operatorname{ms}$ for typical parameters \cite{temporalcorr}. Note that this implies that the disturbance of the condensate in each measurement is negligible. The sequential interactions can be achieved using the dependence of the coupling strength on the value of the magnetic field near a Feshbach resonance. For the dot that is interacting with the condensate, the value of $g_{ab}$ is slightly different from $g_{bb}$, giving rise to a finite value of $g$. For the rest of the dots, $g_{ab}=g_{bb}$ and $g=0$. When $g=\delta=0$, as discussed in \cite{recati}, an extra term in the Hamiltonian -which is negligible when the interaction is on- becomes relevant. This term does not generate transitions in the qubit basis, since it is just a dephasing interaction. The ratio between the strength of this dephasing and the dot frequency is $\sqrt{\frac{\hbar k}{8 \rho_b  c V m} }\simeq 10^{-4}$, meaning that decoherence of the dots can be neglected and they can safely be assumed to remain in the initial superposition state while the other dots are being sequentially let to interact and measured

Another way to ensure that different measurements are not correlated is to couple each dot to a different mode of the field. In this case, the interactions can be simultaneous, since each dot would probe a different thermal distribution with the same temperature.  Alternatively, an array of dots with different gaps could be used, for instance taking advantage of a non-homogenous magnetic field. Other sources of imperfections- e.g. laser noise- would also change the details of the dot-condensate interaction -in this case, the coupling strength-, ensuring the statistical independence of the measurements. The only limitation to the number of probes is that the spatial separation must be larger than the healing length of the condensate in order to ensure no direct interaction between them \cite{Orth2008}. Moreover, indirect interaction due to phonon exchange is proportional to $g^2/\delta$, and therefore is negligible for times $t=1/\delta$ and $g/\delta<<1$. Taking for instance a realistic separation between the dots of $300\, \operatorname{nm}$ and a condensate length of $500\,\operatorname{\mu m}$, we can load 1500 impurities in a 1-D optical lattice. This number would be further increased in 2-D or 3-D setups. Since the lifetime of the condensate can be larger than $100 \operatorname{s}$ it seems feasible to perform at least a few measurements on each dot. Therefore, as can be seen in Fig. \ref{fig:results}c) the relative error is well below the 20 $\%$ state of the art thermometry in the sub nK regime \cite{sciencewr}, and ultrahigh precisions below 1 $\%$ can be achieved. Moreover, as explained above, by preparing several dots in an entangled input state \cite{atomchip,wideraentanglement} the precision would be significantly improved reaching the Heisenberg limit. The initial entangled state can be prepared by using a value of the magnetic field that strongly couples the dots to the field \cite{vacengbec}. Then the magnetic field can be varied to decouple the dots and start the experiment, as explained above. Notice that, once the dots are decoupled, the local dephasing interaction cannot change the nonlocal correlations i. e. entanglement.

\section*{Discussion}

Our minimally invasive technique is suitable to measure with high accuracy the very low temperature effects generated in analogue gravity scenarios. The Unruh temperature can in principle be measured in our setting by accelerating the dots. The idea of demonstrating the Unruh effect in BECs was suggested with a different setup in \cite{ciracreznik}. Moreover, the use of a geometric quantum phase  as an estimator of the Unruh temperature was proposed using a challenging experimental setup involving atom Mach-Zehnder interferometry \cite{ivyeduunruhbery}. We are currently working on implementing our technique to measure the Unruh effect in a BEC with higher precision. If $a$ is the acceleration, the Unruh temperature is given by $T_U=\hbar\,a/(2\pi\,c\,k_B)$ which in our setup give rise to $T_U=2.4 \operatorname{nK}$ for an acceleration of $g$ -where $g$ is the acceleration of gravity on the surface of the Earth. Accelerations of several $g$ were achieved in recent experiments \cite{meschede}. Therefore, with our method the detection of the Unruh temperature of the phononic BEC bath is clearly within reach of current technology.

We have shown that the quantum dynamical phase acquired by a two-level system interacting with a quantum field in a thermal state under the Jaynes-Cummnigs Hamiltonian can be used to measure the temperature of the field. Since the Jaynes-Cummings model successfully describes a number of different quantum systems, our ideas can be applied to measure the temperature in a great variety of physical setups, such as cavity QED, ion traps or circuit QED in a weak coupling regime. As a specially interesting implementation we choose to present in this paper a cold atom setup. In particular, an atomic quantum dot is coupled to the phononic quantum fluctuations of a BEC.  We compute the dynamical phase acquired by several independent dots in a Ramsey interferometry scheme and estimate the precision by means of the Quantum Fisher Information. We show that the phase is sensitive to temperatures in the sub nK regime and that the precision improves on the state of the art thermometry at those temperatures after a number of measurements that can be achieved with current technology. Moreover, the precision can in principle be further improved by preparing entangled input states. 

The proposal to infer the temperature of the condensate from the phase acquired by an interacting probe, is not restricted to the particular BEC model we present here.  We chose as a particular example a probe coupled to a single mode of a weekly interacting BEC because of its mathematical simplicity and elegance.  In principle, our ideas can also be applied to different cold atom systems, such as Fermi gases -where they would be useful in the analysis of magnetic phase transitions. Moreover, other states of the field might be considered and therefore the readout of the qubit's phase could be a tool to analyse the dynamics of a BEC and other cold-atom systems.

In summary, we have introduced a non-invasive, accurate and quantum way of measuring ultralow temperatures in a Bose-Einstein Condensate. In addition to improving on state-of the art thermometry in the sub nK regime, these techniques provide an experimentally feasible method of measuring very-low temperatures arising in analogue gravity scenarios, paving the way for the detection of analogue Unruh and Hawking temperatures in BECs.   

\section{Methods}

\subsection{Dynamical and geometrical phases}

The time evolution of the system can be found by  writing  $H_{JC} (t)= U(\theta)H^0_{JC}U(\theta)^{\dagger}$ where $U(\theta)=e^{i\,\theta\,a^{\dagger}a}$ and $H^0_{JC}=H_{JC}(x=0,t=0)$. The eigenstates of $H^0_{JC}$, are given by $\ket{n-}=\operatorname{cos}(\frac{\alpha_n}{2})\ket{(n+1)g}-\operatorname{sin}(\frac{\alpha_n}{2})\ket{ne}$ and $\ket{n+}=\operatorname{sin}(\frac{\alpha_n}{2})\ket{(n+1)g}+\operatorname{cos}(\frac{\alpha_n}{2})\ket{ne}$  where $\ket{n\,g}$ and $\ket{n\,e}$ are the eigenstates of the free Hamiltonian (corresponding to $g=0$) and $\alpha_n=\operatorname{arctan}(\frac{2\,g\,\sqrt{n+1}}{\delta})$.
In this notation $|n\rangle$ are field number states and $\ket{g}$ and $\ket{e}$ are the ground and excited states of the qubit. We require that the qubit only acquires phases during its evolution. Therefore, to ensure that there are no transitions between the $U(\theta)\ket{n\pm}$ states, we make use of the adiabatic approximation  \cite {ballentine} $\sum_m\frac{|\bra{m}\dot{H}\ket{n}|}{E_m-E_n}\, t<<1$. In the Jaynes-Cummings model transitions are only possible between the eigenstates with same $n$, therefore the adiabatic condition yields $\frac{|\bra{n+}\dot{H}\ket{n-}|}{E_{n+}-E_{n-}} t \simeq  \frac{gt}{2} << 1 $. Notice that with the values of $g$ and $t$ that we consider in the text the adiabatic condition  holds.
For simplicity, we assume that the parameter $\theta$ undergoes a cycle, therefore the interaction time is $t= 2\pi/\delta$. The dynamical phase $\gamma_{Dn}^{\pm}$ acquired by the state $\ket{n\pm}$ under this condition is given by 
\begin{eqnarray}
\gamma_{Dn}^{\pm}&=&-\frac{1}{\hbar}\int_0^t\, <H>_{n\pm}d t'=-\frac{E_n^{\pm}\,t}{\hbar} =\nonumber\\&=&-\frac{2\pi}{\delta}(\Omega_a (n-\frac{1}{2})\pm
\operatorname{\sqrt{\delta^2+4\,g^2\,n}}). \label{eq:gammasd}
\end{eqnarray}
The geometric phase $\gamma_{Gn}^{\pm}$ defined by  \cite{ivyvlatkoberry}: $i\gamma_{Gn}^{\pm}=\int_0^{2\pi}\, <U^{\dagger}\partial_{\theta}U>_{n\pm}d \theta$, yields $\gamma_{Gn}^{+}=2\pi (n-\operatorname{cos}^2(\frac{\alpha_n}{2}))$ and $\gamma_{Gn}^{-}=2\pi (n-\operatorname{sin}^2(\frac{\alpha_n}{2}))$.
We are interested in the case where the field is initially in a thermal state $\rho_F=\sum_n p_n\ket{n_f}\bra{n_f}$, where $p_n=(e^{-F})^n\,(1-e^{-F})$, $F=\frac{\hbar\Omega_a}{k_B\,T}$.  In the mixed case, the geometric phase is given by $\gamma_G^{\pm}=\operatorname{Arg}(\sum_n p_n\,e^{(i\gamma^{\pm}_{Gn})})$ \cite{geometricmixed1,geometricmixed2}, while in the dynamical case, the phase is obtained through a simpler expression $\gamma_D^{\pm}=\sum_n\,p_n\gamma_{Dn}^{\pm}$. 
We find, using a series expansion on $g/\delta$, that the geometric phase is given by Equation (\ref{eq:geometricalappr}) and the dynamical phase by Equation (\ref{eq:dynamicalappr})
\subsection{Quantum Fisher Information}

The quantum Cramer-Rao bound \cite{advances} states that the error $\delta x$ in estimating a parameter $x$ with $M$ measurements on a state $\rho_x$ that depends on $x$, is bounded by: 
\begin{equation}
\delta x\geq\frac{1}{\sqrt{M\,F(\rho_x)}},\label{eq:error}
\end{equation}
$F(\rho_x)$ being the Quantum Fisher Information of the state.
This inequality assumes that input states are separable, however, it has been shown that entangled states improve the $1/\sqrt{M}$ shot-noise scaling reaching the Heisenberg limit given by a $1/M$ scaling where $M$ is the number of entangled particles.
$F(\rho_x)$ can be written as  \cite{parisfisher}:
\begin{equation}
F(\rho_x)= 2 \sum_{nm}\frac{|\bra{m}\partial_x\rho_x\ket{n}|^2}{\rho_m+\rho_n}\label{eq:fisher}
\end{equation}
where $\rho_{m,n}$ are the matrix elements of $\rho_x$ in an orthogonal basis in which $\rho_x$ is diagonal: $\rho_x=\sum_n \rho_n \ket{n}\bra{n}$. The terms in which both $\rho_m$ and $\rho_n$ are zero in the sum in Eq. (\ref{eq:fisher})  must be excluded. 

\section*{Acknowledgements} 
We are indebted to Tomi Johnson for his very helpful insights. The authors would like also to thank Gerardo Adesso, Mehdi Ahmadi, Michael Berry, Per Delsing, G{\"o}ran Johansson, Tim Ralph, Enrique Solano and Daniel Oi for useful discussions and the financial support of EPSRC Bridging the gaps.  C.~S and I.~F. acknowledge support from EPSRC (CAF Grant No.~EP/G00496X/2 to I.~F.) A.~W. acknowledges funding from EPSRC grant No. EP/H027777/1.


\end{document}